\documentclass[twocolumn,amssymb,prb,showpacs]{revtex4}
\usepackage{epsfig}
\usepackage{graphicx}

\begin{document}
\title{Tunnelling magnetoresistance anomalies of a Coulomb blockaded quantum dot}

\author{Piotr Stefa\'nski}
\email{piotrs@ifmpan.poznan.pl}
 \affiliation{Institute of Molecular Physics of the Polish Academy of Sciences\\
  ul.~Smoluchowskiego 17, 60-179 Pozna\'n, Poland}
\author{}
\affiliation{}
\author{}
\affiliation{}
\date{\today}

\begin{abstract}
We consider quantum transport and tunneling magnetoresistance
(TMR) through an interacting quantum dot in the Coulomb blockade
regime, attached to ferromagnetic leads. We show that there exist
two kinds of anomalies of TMR, which have different origin. One
type, associated with the TMR sign change and appearing at
conductance resonances, is of a single particle origin. The second
type, inducing  a pronounced increase of TMR value far beyond 100
\%, is caused by electron correlations. It is manifested
in-between Coulomb blockade conductance peaks. Both the types of
anomalies are discussed for zero and finite bias and their
robustness to the temperature increase is also demonstrated. The
results are presented in the context of recent experiments on
semiconductor quantum dots in which similar features of TMR have
been observed.
\end{abstract}
\maketitle

\section{introduction}
Spin dependent tunnelling phenomena have attracted much scientific
attention recently mostly due to the promising potential
applications for magnetic sensors and magnetic random access
memories. Early work of Moodera \textit{et. al.}\cite{moodera},
showing large, reproducible TMR effect of ferromagnetic tunnel
junctions with $Al_{2}0_{3}$ spacer measured at room temperature,
was one of the first steps initiating an avalanche of both
theoretical and experimental investigations. The results were well
understood within Julliere's model \cite{julliere}. Within this
model the TMR effect is described in terms of the densities of
states polarizations at Fermi energy of left (L) and right (R)
ferromagnetic electrodes: $TMR=2P_{L}P_{R}/(1-P_{L}P_{R})$. The
electronic structure of the spacer is assumed to be featureless.
It appeared that this model was too simple to understand the
experimental results that further arose. The interface resonant
states appearing as a result of energy structure mismatch of the
ferromagnetic lead and insulating spacer have considerable
influence of the TMR value \cite{tsymbal,zutic}. In particular,
symmetry of these states can select spin-polarized bands in
ferromagnetic electrode and enhance tunnelling through the
insulating barrier. Moreover, a modification of the energy
structure of the spacer and its bonding to the ferromagnetic leads
can switch the polarization of the current and also change the
sign of tunnelling magnetoresistance. These features can be
controlled by a proper modification of the spacer composition
\cite{tsymbal}. It was also shown that the defect states, present
in the insulating spacer separating two ferromagnetic leads, can
dramatically change the exchange interaction between the leads
\cite{zhuravlev} and affects TMR in the way not understood within
Julliere's model. Recently it has also been derived that the
scattering of electrons on nonmagnetic impurities present in the
nanojunction barrier can cause the switching of TMR sign
\cite{ren}.

In recent years, due to a rapid development of nanotechnology, a
new kind of "spacer" became available to put between ferromagnetic
leads. Importantly, the electronic properties of these "spacers"
are well controllable. These are semiconductor quantum
dots\cite{jacak} (QDs). They have well defined discrete energy
spectrum, whose position with respect to the chemical potential of
the leads can easily be capacitatively tuned by electric field of
a nearby gate. This in turn gives the possibility of electrical
control of the TMR effect for such a device, which opens new
possibilities of applications. Manipulation of the spin by
electric field is one of the central issues of spintronics, which
is regarded as a promising alternative for traditional
charge-based electronics. Electron interactions inside the dots
cause dramatic effects in their conductance, displaying quantum
Coulomb blockade \cite{kouwenhoven} and Kondo effect
\cite{goldhaber}. Recently TMR measurements have been performed
for such designed devices: InAs quantum dots coupled to nickel or
cobalt electrodes \cite{hamaya1,hamaya2,hamaya3,igarashi}. These
experiments show a rich TMR behavior, including the changes of the
TMR sign and appearance of its maximum far exceeding 100 \%. Spin
transport and gate control of the tunnelling magnetoresistance has
also been realized in carbon nanotubes \cite{saho, jensen}. The
variety of TMR anomalies observed in those systems is also
ascribed to the discreetness of the nanotube energy structure.

Tunnelling magnetoresistance oscillations caused by the classical
Coulomb blockade at a small metallic droplet coupled to
ferromagnetic electrodes had already been predicted  long
ago\cite{barnas}. The TMR value can also change its sign due to
strong electron correlations inside the dot in Kondo regime, as
was shown theoretically \cite{zhang,choi}. It is a result of the
current enhancement by the Kondo resonance when the dot is coupled
to the leads of antiparallel spin configuration.

In the present paper we make an attempt to describe experimentally
encountered anomalies of TMR  \cite{hamaya1,hamaya2,hamaya3} for
the InAs quantum dots in Coulomb blockade regime. We utilize the
model of an interacting quantum dot with one level active in
transport,
 coupled to ferromagnetic leads. We show that TMR sign
 switching is caused by the resonances of the dot level with one
of the leads chemical potential in presence of the large asymmetry
of the dot-leads coupling. We also predict that electron-electron
interactions inside the dot have a decisive role in the formation
the TMR maximum, which exceeds  100 \%  . Recently, within the
similar model device, we also introduced a proposal of the
electrical control of the spin polarization of the current
\cite{firstpaper}. We will show that the correlation induced
switching of the spin direction of the current is closely related
to the TMR sign change at Coulomb blockade.

\section{Theoretical approach}
The device is described by Anderson hamiltonian \cite{anderson61},
where the dot takes the role of magnetic impurity and the
(polarized) leads are analogues of host metal:
\begin{eqnarray}
\label{for1}  \nonumber H= \epsilon_{d}d_{\sigma}^{+}d_{\sigma}+U
 n_{\sigma}n_{\bar{\sigma}}
 +\sum_{k,\sigma,\alpha=L,R}\lbrack
 t_{\alpha}c_{k\alpha,\sigma}^{+}d_{\sigma}+h.c.]\\+
\sum_{k,\sigma,\alpha=L,R}\epsilon_{k\alpha,\sigma}c_{k\alpha,\sigma}^{+}c_{k\alpha,\sigma}
\end{eqnarray}
The first two terms describe the dot with the presence of Coulomb
interactions $U$. The bare dot level is shifted
 by the gate voltage acting on the dot capacitatively:
$\epsilon_{d}\equiv\epsilon_{d}-V_{g}$, and its initial position
for $V_{g}=0$ is assumed to coincide with Fermi level
$\epsilon_{d}=\epsilon_{F}=0$. The third term describes the
tunnelling between the dot and the leads, represented by the last
term in Eq.~(\ref{for1}). The electron energy in the leads is
spin-dependent, $\sigma=\uparrow,\downarrow$, because the leads
are assumed to be spin polarized. We neglect the spin dependence
of the tunnelling matrix elements $t_{\alpha}$ ($\alpha=L, R$)
which are rather dependent on the potential barrier between the
dot and a given lead. Thus, the spin dependence of the QD level
width
$(\Gamma_{\sigma}/2)=(1/2)\sum_{\alpha}\Gamma_{\alpha\sigma}$;
$\Gamma_{\alpha\sigma}=2\pi|t_{\alpha}|^{2}\rho_{\alpha\sigma}$ is
caused by the coupling to the leads with different spectral
densities $\rho_{\alpha\uparrow}\neq\rho_{\alpha\downarrow}$,
which are assumed to be featureless and constant.

Let us define the polarization of the quantity $X$,
$P_{X}=(X_{\uparrow}-X_{\downarrow})/(X_{\uparrow}+X_{\downarrow})$.
For the lead $\alpha$ it is:
$P_{\alpha}=(\rho_{\alpha\uparrow}-\rho_{\alpha\downarrow})/(\rho_{\alpha\uparrow}+\rho_{\alpha\downarrow})$,
which can be expressed by the spin-dependent QD widths:
\begin{equation}
\label{lead_pol}
P_{\alpha}=(\Gamma_{\alpha\uparrow}-\Gamma_{\alpha\downarrow})/(\Gamma_{\alpha\uparrow}+\Gamma_{\alpha\downarrow}).
\end{equation}
To calculate TMR we will consider parallel (P), $P_{R}=P_{L}$, and
antiparallel (AP), $P_{R}=-P_{L}$, leads polarizations arrangement
 The asymmetry of the dot-leads coupling is
described by $\alpha$ parameter. Because of validity of
Eq.~(\ref{lead_pol}), the relations between dot level width
components from left and right lead follow for (P) and (AP)
configurations:
\begin{equation}
\label{gamma_rel}
 \Gamma_{R\sigma}^{P}=\alpha\Gamma_{L\sigma}^{P},
\quad
\Gamma_{R\sigma}^{AP}=\alpha\Gamma_{L\bar{\sigma}}^{AP},\quad
\sigma,\bar{\sigma}=\uparrow, \downarrow.
\end{equation}
Tunnelling magnetoresistance is calculated from the formula:
$TMR=(\mathcal{G}^{P}-\mathcal{G}^{AP})/\mathcal{G}^{AP}$, where
$\mathcal{G}$ are appropriate conductances calculated for parallel
and antiparallel configurations.

The retarded dot Green function
$G_{\sigma}^{r}(t-t')=-i\theta(t-t')\langle
d_{\sigma}(t)d_{\sigma}^{\dagger}(t')+d_{\sigma}^{\dagger}(t')d_{\sigma}(t)\rangle$
is obtained by solving the set of equations of motion of the Green
functions in the Hubbard I approximation \cite{hewson}. Within
this approximation the two-particle Green functions describing
spin-flip processes (generating Kondo effect) on the localized
level are neglected. The Green functions that describe the normal
scattering of band electrons on an impurity are approximated by
decoupling of band electrons from impurity electrons. The Hubbard
approximation is valid for large $U/\Gamma$ ratio, when the
Hubbard subbands are well separated in energy scale. For numerical
calculations we assumed $\Gamma_{L\uparrow}=0.3U$, the other width
components are calculated from Eqs.~(\ref{lead_pol}) and
(\ref{gamma_rel}) for given lead polarization and  asymmetry
$\alpha$.

The Hubbard approximation is the simplest scheme which describes
correlated electrons, placed on the approximation scale between
Hartree-Fock approximation for interacting but uncorrelated
electrons, and the schemes for strongly correlated electrons,
leading to Kondo physics. Thus, it is most suitable for the
description of a spin-degenerate QD level in the Coulomb blockade
regime of the lead-dot coupling, the limit realized in recent
experiments\cite{hamaya1,hamaya2,hamaya3}.

The Fourier-transformed expression for QD Green function with the
spin $\sigma=\uparrow,\downarrow$ for given $\beta=P$ or $AP$
arrangement has the form:
\begin{eqnarray}
\label{G1gen} \nonumber
G_{\sigma}^{r,\beta}(\omega)=[\frac{\omega-\epsilon_{d}}
{1+\frac{\langle n_{\bar{\sigma}}\rangle^{\beta}
U}{\omega-\epsilon_{d}-U}}+\frac{i\Gamma_{\sigma}^{\beta}}{2}]^{-1}\\\simeq
\frac{1-\langle
n_{\bar{\sigma}}\rangle^{\beta}}{\omega-\epsilon_{d}+\frac{i\Gamma_{\sigma}^{\beta}}{2}}+\frac{\langle
n_{\bar{\sigma}}\rangle^{\beta}}{\omega-\epsilon_{d}-U+\frac{i\Gamma_{\sigma}^{\beta}}{2}}.
\end{eqnarray}

Eq. (\ref{G1gen}) has been written as the sum of two hubbard
resonances, $\epsilon_{d}^{I}=\epsilon_{d}$ and
$\epsilon_{d}^{II}=\epsilon_{d}+U$, whose spectral weights are
controlled by the dot level occupancy with the opposite spin
$\bar{\sigma}$. This feature directly reflects  Coulomb
interactions between electrons with opposite spins.

The spin components of the dot occupancy  have been calculated
selfconsistently from the set of coupled equations:
\begin{eqnarray}
\label{self}
 \nonumber \langle n_{\sigma}\rangle^{\beta} =-\frac{i}{2\pi}\int
G^{<,{\beta}}_{\sigma}(\omega,\langle n_{\bar{\sigma}}\rangle^{\beta})d\omega,\\
\langle n_{\bar{\sigma}}\rangle^{\beta} =-\frac{i}{2\pi}\int
G^{<,{\beta}}_{\bar{\sigma}}(\omega,\langle
n_{\sigma}\rangle^{\beta})d\omega.
\end{eqnarray}
The "lesser" dot Green function $G^{<,\beta}$ can be expressed by
the spectral density of the dot \cite{haug},
$\rho_{\sigma}^{\beta}(\omega)=-(1/\pi)\Im
G^{r,\beta}_{\sigma}(\omega)$,
$G^{<,\beta}_{\sigma}(\omega)=2i\pi\bar{f}(\omega)\rho_{\sigma}^{\beta}(\omega)$.
Non-equilibrium distribution function
$\bar{f}=[\Gamma_{L\sigma}^{\beta}f_{L}+\Gamma_{R\sigma}^{\beta}
f_{R}]/(\Gamma_{L\sigma}^{\beta}+\Gamma_{R\sigma}^{\beta})$ has a
two-step profile defined by the chemical potential in the leads:
$f_{L/R}\equiv f(\omega\mp eV)$ and collapses into equilibrium
Fermi-Dirac distribution function $f\equiv f_{L}=f_{R}$ in the
limit of zero bias between the leads, $eV\rightarrow 0$.
The current is calculated within Landauer formalism from the
relation \cite{haug}:
\begin{equation}
J^{\beta}=\frac{e}{2\hbar}\sum_{\sigma}\int d\omega
[f_{L}-f_{R}]\frac{\Gamma_{L\sigma}^{\beta}\Gamma_{R\sigma}^{\beta}}{\Gamma_{L\sigma}^{\beta}+\Gamma_{R\sigma}^{\beta}}\rho_{\sigma}^{\beta}(\omega).
\end{equation}
In the limit of zero bias the conductance has the form:
\begin{equation}
\mathcal{G}^{\beta}=\frac{\partial J^{\beta}}{\partial V}
=\frac{e^2}{\hbar}\sum_{\sigma}\int d\omega (-\frac{\partial
f}{\partial
\omega})\frac{\Gamma_{L\sigma}^{\beta}\Gamma_{R\sigma}^{\beta}}{\Gamma_{L\sigma}^{\beta}+\Gamma_{R\sigma}^{\beta}}\rho_{\sigma}^{\beta}(\omega).
\end{equation}

\section{Behavior of the system at zero bias}
\begin{figure} [tr]
\epsfxsize=0.48\textwidth \epsfbox{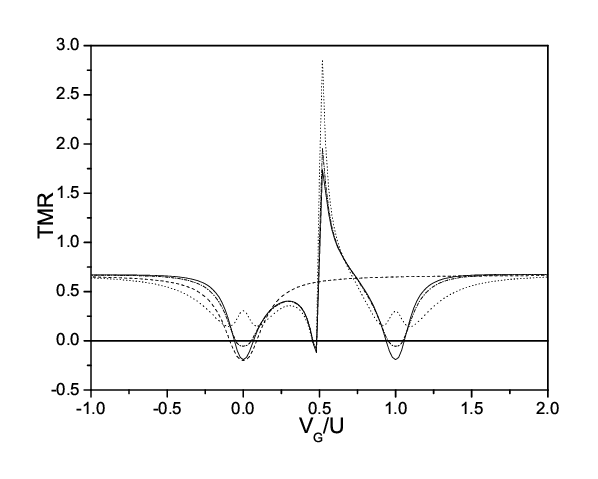} \caption{\label{fig1}
TMR dependence on gate voltage calculated for $T=0.01U$,
$P_{L}=0.5$ and zero bias for various asymmetry parameters
:$\alpha=0.1$- solid, $\alpha=1$-dotted, $\alpha=0.2$ dash-dotted
curve. The dashed line is for non-interacting dot and
$\alpha=0.1$.}
\end{figure}
For all numerical results presented we have chosen the left lead
polarization $P_{L}=0.5$ and temperature $T=0.01U$. This range  is
typical for experiments \cite{hamaya1,hamaya2,hamaya3}; it gives
the temperature of $174 mK$ for $U=15 meV$ \cite{hamaya2}.

In Fig.~(\ref{fig1}) calculated TMR evolution with the change of
gate voltage for different $\alpha$ asymmetry parameter is
demonstrated. In order to understand various TMR anomalies shown
in Fig.~(\ref{fig1}) it is instructive to analyze analytically TMR
expression at $T=0$. The conductance for a given spin and leads
configuration $\beta$ has the following form in this limit:
\begin{equation}
\label{G_sig}
\mathcal{G}_{\sigma}^{\beta}=\frac{e^2}{h}\frac{\Gamma_{L\sigma}^{\beta}\Gamma_{R\sigma}^{\beta}}{[\frac{\epsilon_{d}(\epsilon_{d}+U)}{\epsilon_{d}+U(1-\langle
n_{\bar{\sigma}}\rangle^{\beta})}]^2+\frac{1}{4}(\Gamma_{L\sigma}^{\beta}+\Gamma_{R\sigma}^{\beta})^2}
\end{equation}
Consider the situation when $\epsilon_{d}^{I}$ or
$\epsilon_{d}^{II}$ hubbard level crosses Fermi level (at
$V_{g}=0$ or $V_{g}=U$). In this case Eq.~(\ref{G_sig}) takes the
form:
\begin{equation}
\mathcal{G}^{\beta}_{\sigma}=\frac{e^2}{h}\frac{\Gamma_{L\sigma}^{\beta}\Gamma_{R\sigma}^{\beta}}{\frac{1}{4}(\Gamma_{L\sigma}^{\beta}+\Gamma_{R\sigma}^{\beta})^2}.
\end{equation}
The conductance has exactly the same form as for non-interacting
dot level of Green's function:
$\mathcal{G}^{r}_{\sigma}=[\omega-\epsilon_{d}+i(\Gamma_{\sigma}^{\beta}/2)]^{-1}$
crossing Fermi level, $\epsilon_{d}=\epsilon_{F}$. Thus, these TMR
features are of the single particle origin and can be described in
the limit of non-interacting electrons. In Fig.~(\ref{fig1}) also
the TMR curve for non-interacting dot level, calculated for
$\alpha=0.1$, is shown. The minimum at $V_{g}=0$ coincides with
the one of interacting case for first hubbard level in resonance
with Fermi level $\epsilon_{d}^{I}=\epsilon_{F}$.

Taking into account the relations between the level widths,
Eq.~(\ref{gamma_rel}), it can be shown by straightforward
calculation that for $\epsilon_{d}^{I}=0$ or $\epsilon_{d}^{II}=0$
the spin components of conductance for parallel and antiparallel
configurations are (in units of $e^2/h$):
\begin{equation}
\mathcal{G}^{P}_{\uparrow}=\mathcal{G}^{P}_{\downarrow}=\frac{4\alpha}{(1+\alpha)^2},
\end{equation}
and
\begin{equation}
\label{G_AP}
\mathcal{G}^{AP}_{\uparrow/\downarrow}=-\frac{4\alpha(P_{L}^2-1)}{[P_{L}(1-\alpha)\pm(1+\alpha)]^2},
\end{equation}
where also the relation between spin-dependent widths from
Eq.~(\ref{lead_pol}):
$\Gamma_{L\downarrow}=-\Gamma_{L\uparrow}(P_{L}-1)/(P_{L}+1)$ has
been applied.

For symmetric coupling, $\alpha=1$, the conductance for parallel
configuration
$\mathcal{G}^{P}=\mathcal{G}^{P}_{\uparrow}+\mathcal{G}^{P}_{\downarrow}$
reaches its maximum value of $2e^2/h$. It is reflected in the TMR
curve in Fig.~(\ref{fig1}), which also displays a local maximum at
the conductance resonances at $V_{g}=0$ and $V_{g}=U$. The
conductances for antiparallel configuration and $\alpha=1$ are:
$\mathcal{G}^{AP}_{\uparrow}=\mathcal{G}^{AP}_{\downarrow}=-(P_{L}^2-1)$,
which yields $TMR=P_{L}^2/(1-P_{L}^2)>0$. Thus, for symmetric
dot-leads coupling the TMR has positive sign at the conductance
resonances. The situation changes for asymmetric coupling, $\alpha
< 1$. In this case TMR at resonances takes the form:
\begin{equation}
TMR=\frac{-[P_{L}^2(1-\alpha)^2-(1+\alpha)^2]^2}{(1+\alpha)^2(P_{L}^2-1)[P_{L}^2(1-\alpha)^2+(1+\alpha)^2]}-1,
\end{equation}
and for $\alpha\ll 1$ it changes the sign to negative:
$TMR=-2P_{L}^2/(P_{L}^2+1)<0$, as shown in Fig.~(\ref{fig1}).

Let us summarize the above discussed  single particle TMR sign
changes within a simple physical picture. For the perfect
symmetric coupling, $\alpha=1$, the transmission through the dot
in parallel configuration reaches the conductance quantum in the
both spin channels,
$\mathcal{G}^{P}_{\uparrow}=\mathcal{G}^{P}_{\downarrow}=e^2/h$.
This is caused by the perfect matching of the spectral densities
of spin up and spin down of the left lead to the corresponding
spectral densities of the right lead at the Fermi level. It also
results that the spin dependent level widths due to the coupling
to the left lead and to the right lead are equal, see
Eq.~(\ref{gamma_rel}). In antiparallel configuration and
$\alpha=1$ this is not the case, the numbers of states of spin up
and spin down in the left lead at the Fermi level are different as
compared to the right lead and also the corresponding level widths
are not equal. Thus, the conductance in antiparallel configuration
is less than conductance quantum,
$\mathcal{G}^{AP}_{\uparrow}=\mathcal{G}^{AP}_{\downarrow}=(3/4)(e^2/h)$
for $P_{L}=0.5$, and $TMR > 0$. Note that for unpolarized leads,
$P_{L}=0$, the matching of the spectral densities is retained and
the conductances
$\mathcal{G}^{AP}_{\uparrow}=\mathcal{G}^{AP}_{\downarrow}=e^2/h$
reach conductance quantum. When the asymmetry of the dot-leads
coupling is increased, $\alpha < 1$, the tunnelling between the
dot and the right lead is reduced, which destroys the perfect
matching of the widths in (P) configuration:
$\Gamma^{P}_{R\uparrow}\neq\Gamma^{P}_{L\uparrow}$ and
$\Gamma^{P}_{R\downarrow}\neq\Gamma^{P}_{L\downarrow}$. It results
in a gradual decrease of the both spin conductance components in
(P) configuration with the increase of $\alpha$. More interesting
situation takes place in (AP) configuration. The
$\mathcal{G}^{AP}_{\uparrow}$ conductance component, which
describes the tunnelling of spin up excess electrons ($P_{L} > 0$)
from the left lead via resonant dot state into minority up spin
subband of the right lead ($P_{R} < 0$) decreases with the
increase of the coupling asymmetry because the width
$\Gamma_{R\uparrow}^{AP}$ which is less than
$\Gamma_{L\uparrow}^{AP}$ even for $\alpha=1$ is further decreased
by lowering $\alpha$. It increases the mismatch between the widths
in this spin sector. Different relation is encountered between the
widths  in the (AP) spin down channel. Initially, for $\alpha=1$
we have $\Gamma_{R\downarrow}^{AP}>\Gamma_{L\downarrow}^{AP}$ and
by the decrease of $\alpha$ the value of
$\Gamma_{R\downarrow}^{AP}$ is lowered. When it reaches the value
$\Gamma_{R\downarrow}^{AP}=\Gamma_{L\downarrow}^{AP}$ we have a
perfect symmetric coupling in this channel. Thus, the
\textit{decrease} of the dot-leads coupling symmetry  causes an
\textit{increase} of the coupling symmetry  in the spin down
sector when the system in (AP) configuration. The component
$\mathcal{G}^{AP}_{\downarrow}$ increases and starts to dominate
over other conductance components causing $TMR < 0$ for small
$\alpha$. It can be checked from Eq.~(\ref{G_AP}) that
$\mathcal{G}^{AP}_{\downarrow}$ reaches full transmission limit
for $\alpha=1/3$ when $P_{L}=0.5$. This enhancement of the one of
the conductance spin components in (AP) configuration can easily
be generalized to other (AP) arrangements. For instance, if the
polarization of the left lead were assumed to be negative, the
$\mathcal{G}^{AP}_{\uparrow}$ would be enhanced by the decrease of
$\alpha$.

Similar as discussed TMR sign changes have been observed in
quantum dots \cite{hamaya2,hamaya3} and in carbon nanotubes
\cite{saho} coupled to ferromagnetic electrodes. They are also
interpreted in terms of the asymmetry of the coupling to the
leads.

\begin{figure} [tr]
\epsfxsize=0.45\textwidth \epsfbox{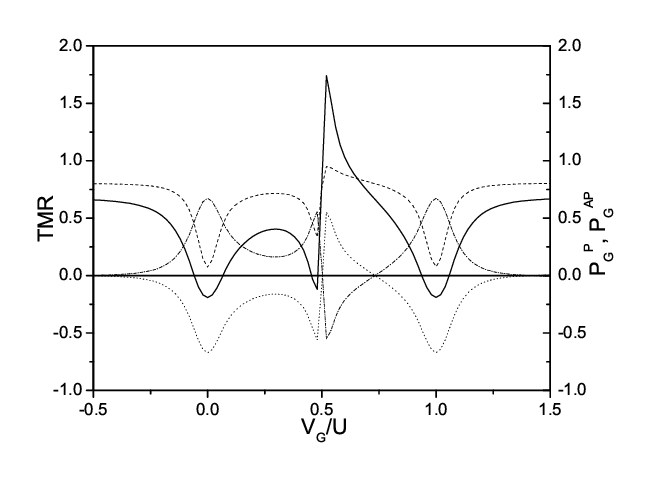}
\caption{\label{fig2}TMR (solid curve) vs. gate voltage calculated
for $\alpha=0.1$ at $T=0.01U$ and zero bias. The corresponding
conductance polarizations $P^{P}_{\mathcal{G}}$ (dashed curve) and
$P^{AP}_{\mathcal{G}}$ (dotted curve) are also shown. Dash-dotted
curve is for $-P^{AP}_{\mathcal{G}}$. }
\end{figure}

Consider now the region in which the dot is in Coulomb blockade.
In this case for  $\epsilon_{d}=-U/2$ exactly one electron is
present at the dot $\langle n_{\uparrow}\rangle^{\beta}+\langle
n_{\downarrow}\rangle^{\beta}=1$. There are considerable TMR
anomalies near this point (see Fig.~(\ref{fig1})), TMR can change
its sign or can be greatly enhanced exceeding 100 percent. These
anomalies are caused by electron-electron interactions . The
peculiar behavior of TMR in this region is caused by interplay of
the two factors. The first is caused by the coupling (and its
asymmetry) of the dot to the spin polarized leads. It determines
the widths of the conductance peaks for parallel and antiparallel
configurations. The second factor are electron correlations which
are the strongest  in this region\cite{comment}. They cause the
spin components of the occupancies to be close to one half at
Coulomb blockade. The TMR anomalies here can be understood by
analyzing the spin components of the conductance for given lead
polarization arrangement, Eq.~(\ref{G_sig}). For $\epsilon_{d}\sim
-U/2$ and $\langle n_{\bar{\sigma}}\rangle^{\beta}\sim 0.5$ the
denominator $\epsilon_{d}+U(1-\langle n_{\bar{\sigma}}\rangle)$ is
very small, causing $\mathcal{G}_{\sigma}^{\beta}$ to be small.
Depending on the leads polarizations and the strength of the
Coulomb interactions the components of the conductances defining
TMR get their minimal values for different positions of the dot
level. The same mechanism causes sudden current polarization
switching in the region of $\epsilon_{d}\sim -U/2$ as discussed in
\cite{firstpaper}. Indeed, it is shown in Fig.~(\ref{fig2}) that
for (P) and (AP) configuration the value of conductance
polarization $P_{\mathcal{G}}^{\beta}$ performs a rapid
oscillation and additionally it changes sign for (AP)
configuration .

TMR value in this region is very sensitive to the change of the
conductance in (AP) configuration. Consider first the perfect
symmetric coupling, $\alpha=1$. For $P_{R}=-P_{L}$ the total
widths of the dot level are equal
$\Gamma_{\uparrow}=\Gamma_{\downarrow}$ because of relation
$\Gamma_{L\uparrow}=\Gamma_{R\downarrow}$ and
$\Gamma_{L\downarrow}=\Gamma_{R\uparrow}$, Eq.~(\ref{gamma_rel}).
Thus, in the symmetric (AP) arrangement the dot behaves as if it
were coupled to unpolarized leads. In such a case spin components
of the conductance are equal,
$\mathcal{G}^{AP}_{\uparrow}=\mathcal{G}^{AP}_{\downarrow}$,  in
the whole range of gate voltages and also the occupancies $\langle
n_{\uparrow}\rangle^{AP}=\langle n_{\downarrow}\rangle^{AP}$ are
equal. The QD occupancy curve displays a plateau at $\langle
n_{\uparrow}\rangle=\langle n_{\downarrow}\rangle\sim 0.5$ due to
Coulomb blockade \cite{firstpaper}. Moreover, for particle-hole
symmetric case, $\epsilon_{d}=-U/2$, the occupancies are $\langle
n_{\uparrow}\rangle^{AP}=\langle n_{\downarrow}\rangle^{AP} =
0.5$, giving
$\mathcal{G}_{\uparrow}^{AP}=\mathcal{G}_{\downarrow}^{AP}=0$.
This feature causes an infinite TMR value for $\alpha = 1$ at
$T=0$. For asymmetric coupling, $\alpha < 1$ and finite
temperature, the relation
$\mathcal{G}^{AP}_{\uparrow}=\mathcal{G}^{AP}_{\downarrow}$ is
still fulfilled at the gate voltage of TMR maximum, but the
conductances have small finite value. From the condition of
$\mathcal{G}^{AP}_{\uparrow}=\mathcal{G}^{AP}_{\downarrow}$
follows also the equality  $\langle
n_{\uparrow}\rangle^{AP}=\langle n_{\downarrow}\rangle^{AP}$ at
gate voltage of TMR maximum, independently on $\alpha$. It can
easily be derived utilizing Eq.~(\ref{G_sig}). At Coulomb blockade
the first term in the denominator is much larger than the second
one and the relations Eq.~(\ref{gamma_rel}) for (AP) configuration
have to be used.
 Thus, at TMR
maximum the conductance in (AP) configuration is unpolarized (the
conductance polarization $P_{\mathcal{G}}^{AP}=0$ as shown in
Fig.~(\ref{fig2})) and also the dot occupancy polarization is
zero.

Strong enhancement of TMR at classical Coulomb blockade has also
been predicted for ferromagnetic double tunnel junctions
\cite{takahasi} as a result of cotunneling of electrons through
metallic island. It has also been observed experimentally
\cite{schlep,ohno}. Our result provides an explanation of
corresponding TMR maximum at quantum Coulomb blockade for
semiconductor quantum dot. It is in relation with the recent TMR
measurements for InAs quantum dot coupled to the Ni leads
\cite{hamaya3}, where the TMR enhancement above 300 \% has been
observed at Coulomb blockade.

At the gate voltages for which $TMR=0$ following equalities apply
$\mathcal{G}^{AP}_{\uparrow}=\mathcal{G}^{P}_{\downarrow}$ and
$\mathcal{G}^{AP}_{\downarrow}=\mathcal{G}^{P}_{\uparrow}$. It
implies that the conductance spin polarizations are opposite for
parallel and antiparallel configuration:
$P^{P}_{\mathcal{G}}=-P^{AP}_{\mathcal{G}}$ (compare corresponding
curves in Fig.~(\ref{fig2})).

Let us now discuss the TMR sign change encountered at Coulomb
blockade. It is shown in Fig.~(\ref{fig2}) that this TMR minimum
coincides with the sharp minima of conductance polarizations,
moreover $P_{\mathcal{G}}^{AP}<0$ here. The TMR minimum is caused
by sharp polarization switching of the conductance in (AP)
configuration. It is in contrast to (P) configuration, for which
the conductance polarization does not change its sign. In (P)
configuration the dot is coupled to the leads both having excess
of electrons of the same spin (in our case spin up) and
conductance of this spin dominates in the whole range of gate
voltages, $P^{P}_{\mathcal{G}}>0$ (Fig.~(\ref{fig2})). In (AP)
configuration, on the average, there is no such an excess of
electrons of particular spin coming from the leads. Thus,
$\mathcal{G}^{AP}_{\uparrow}$ and $\mathcal{G}^{AP}_{\downarrow}$
are comparable in magnitude and the interactions between spin up
and spin down electrons cause more dramatic changes in the
conductance polarization. Apart from electron-electron
interactions, which cause the conductance polarization switching
in (AP) configuration, the mechanism of enhancement of the
conductance component $\mathcal{G}^{AP}_{\downarrow}$ for small
$\alpha$ works also here, similarly as discussed for single
particle origin TMR sign change. The transmission in the spin down
channel of (AP) configuration dominates here over other
conductance components and causes $TMR <0$. There is no unique
condition for negative TMR minimum to appear at Coulomb blockade
and it depends on the value of initial leads polarization. For
higher $P_{\alpha}$, the conductance in (P) configuration will
dominate ($P^{P}_{\mathcal{G}}$ curve then is shifted upward) and
TMR will not change sign in spite of conductance polarization
switching in (AP) configuration (not shown). The polarization
$P^{AP}_{\mathcal{G}}$ behaves similarly as in the case of
configuration with one lead polarized \cite{firstpaper}. The TMR
sign change here also fades out rapidly with increase of
temperature, Fig.~(\ref{fig3}), because the $P^{AP}_{\mathcal{G}}$
is very sensitive to the temperature change in this region. At
higher temperatures $P^{AP}_{\mathcal{G}}$ switching decays
quickly. This sensitivity in the region of $V_{g}\sim U/2$ has
been shown in \cite{firstpaper} for one of the leads polarized.

It could be counterintuitive that the conductance polarization
$P_{\mathcal{G}}^{P}$ reaches values larger than the leads
polarizations $P_{\alpha}=0.5$ for $|V_{g}|\ll \epsilon_{F}$, as
displayed in Fig.~(\ref{fig2}). We demonstrate that it is the
case. In this limit the dot level is placed far from Fermi level,
thus the dot is unoccupied or fully occupied. Consider the
unoccupied dot for configuration $\beta$, which is realized for
large negative gate voltage. Inserting $\langle
n_{\bar{\sigma}}\rangle^{\beta}=0$ into Eq.~(\ref{G_sig}) and
noting that $(\epsilon_{d})^2\gg(\Gamma_{\sigma}/2)^2$ the
conductance polarization can be written as:
\begin{equation}
\label{P_Ggammas}
P_{\mathcal{G}}^{\beta}=\frac{\Gamma_{L\uparrow}^{\beta}\Gamma_{R\uparrow^{\beta}}-\Gamma_{L\downarrow}^{\beta}\Gamma_{R\downarrow}^{\beta}}
{\Gamma_{L\uparrow}^{\beta}\Gamma_{R\uparrow}^{\beta}+\Gamma_{L\downarrow}^{\beta}\Gamma_{R\downarrow}^{\beta}}.
\end{equation}
For $\beta=P$ the relations of Eq.~(\ref{gamma_rel}) are further
utilized to give:
\begin{equation}
\label{P_Guu}
P_{\mathcal{G}}^{P}=\frac{\Gamma_{L\uparrow}^2-\Gamma_{L\downarrow}^2}
{\Gamma_{L\uparrow}^2+\Gamma_{L\downarrow}^2}=\frac{2P_{L}}{P_{L}^2+1}.
\end{equation}
The last r.h.s expression has been obtained by substituting
$\Gamma_{L\downarrow}$ calculated from Eq.~(\ref{lead_pol}). Note
that the result is independent on $\alpha$. Thus, for
$P_{L}=P_{R}=0.5$ the conductance polarization
$P_{\mathcal{G}}^{P}=0.8$. The same expression,
Eq.~({\ref{P_Guu}), is obtained for fully occupied dot $\langle
n_{\bar{\sigma}}\rangle^{P}=1$  and noting that
$(\epsilon_{d}+U)^2\gg(\Gamma_{\sigma}/2)^2$.
 For antiparallel
configuration, the conductance polarization
$P_{\mathcal{G}}^{AP}=0$ in the limit of $|V_{g}|\ll
\epsilon_{F}$, as shown in Fig.~(\ref{fig2}). It can
straightforwardly be derived from Eq.~(\ref{P_Ggammas}) for
$\beta=AP$ when Eq.~(\ref{gamma_rel}) is utilized:
\begin{figure} [tr]
\epsfxsize=0.45\textwidth \epsfbox{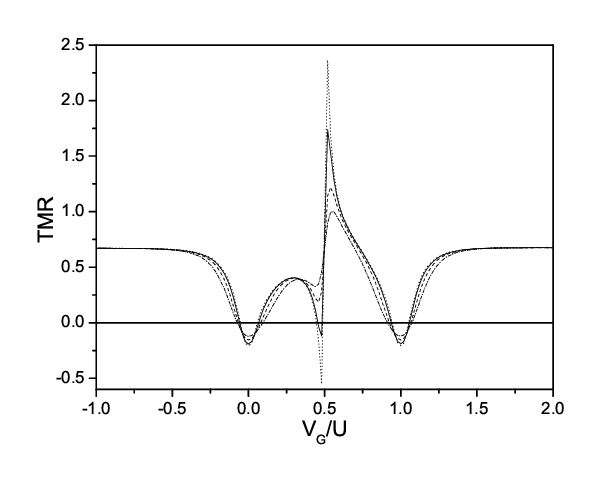}
\caption{\label{fig3}Temperature dependence of the zero bias TMR
calculated for $\alpha=0.1$ and $P_{L}=0.5$: $T=0$-dotted,
$T=0.01U$-solid, $T=0.02U$-dashed and $T=0.03U$-dash-dotted
curve.}
\end{figure}
Note that TMR in the discussed limit of $|\epsilon_{d}|\ll
\epsilon_{F}$ reaches value $2/3$ as predicted by Julli\`{e}re's
model\cite{julliere} for two polarized leads separated by
featureless tunnel barrier.

Temperature dependence of the TMR for asymmetric coupling
$\alpha=0.1$ is shown in Fig.~(3). TMR anomalies in the regions of
$V_{g}\sim 0$ and $V_{g}\sim U$ caused by the resonances of QD
hubbard levels $\epsilon^{I}_{d}$ and $\epsilon^{II}_{d}$ with the
Fermi level are robust to the increase of temperature. The
anomalies due to electron correlations situated in the range of
Coulomb blockade $V_{g}\sim U/2$ in turn, are sensitive to these
changes. It is caused by a temperature increase of the conductance
spin components at the Coulomb blockade valley. However, the
pronounced TMR maximum reaching 100 percent survives for $T=0.03U$
($\simeq 0.5 K$ for $U=15 meV$ \cite{hamaya2}). Contrary, the
negative TMR minimum rapidly disappears at higher temperature, as
pointed out previously.
\section{Behavior of the system at finite bias}
If the finite bias is applied to the system, an electron transport
through excited states of the QD can be activated. In the Hubbard
approximation, used  for description of the device, these
processes are not taken into account.

We consider two limits of the value of the bias applied as
compared to the width of the QD level: small bias limit:
$|eV|<\Gamma_{\sigma}$ and large bias limit when
$|eV|>\Gamma_{\sigma}$. In the large bias limit the chemical
potentials of the leads are well separated in energy scale. It
implies also a good separation of the differential conductance
resonance peaks which appear when any of the hubbard QD levels
crosses given chemical potential. Thus, the maxima and minima of
TMR are also well separated. In the small bias limit the
conductance resonances overlap each other which causes splitting
of TMR minima and diminishing of TMR maximum at Coulomb blockade.
In both the limits, a correspondence can be found between TMR
features with those appearing at zero bias.
\begin{figure} [tr]
\epsfxsize=0.45\textwidth \epsfbox{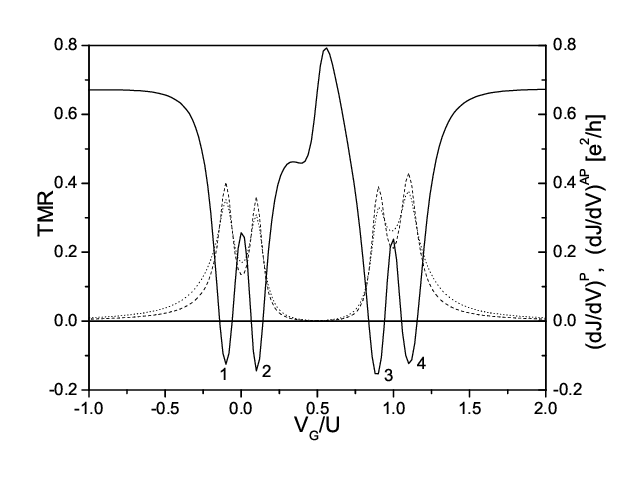}
\caption{\label{fig4}TMR dependence on gate voltage (solid curve)
for finite bias $eV=0.1U$ calculated at $T=0.01U$, $P_{L}=0.5$ and
asymmetric coupling to the leads $\alpha=0.1$. The corresponding
differential conductances $(\partial J/\partial V)^{P}$ (dotted
curve) and $(\partial J/\partial V)^{AP}$ (dashed curve) are also
shown. }
\end{figure}

\subsection{Small bias regime}
The TMR behavior for $eV=0.1U$ is shown in Fig.~(\ref{fig4}) for
asymmetric coupling to the leads $\alpha=0.1$ and temperature
$T=0.01U$. One notices that the pronounced TMR minima present for
$V_{g}=0$ and $V_{g}=U$ are split when a finite bias is applied.
For positive bias the left lead chemical potential $\mu_{L}$ is
shifted upwards by $eV$ and  right lead chemical potential
$\mu_{R}$ is shifted downwards by $eV$ on energy scale. When gate
voltage increases from negative values, it shifts the dot level
from empty state regime towards Fermi bathes inside the leads. At
first the $\epsilon_{d}^{I}$ level comes into resonance with
$\mu_{L}$ and then with $\mu_{R}$ chemical potential. It causes
the appearance of two peaks in conductance with the distance of
doubled bias value between them (shown in Fig.~(\ref{fig4})) and
consequently two minima of TMR are produced, labelled by (1) and
(2). Similarly, in the range of $V_{g}\sim U$ the second hubbard
level $\epsilon_{d}^{II}$ comes into resonance with $\mu_{L}$ and
then with $\mu_{R}$ and minima (3) and (4) appear.

The maximum of TMR at $V_{g}\sim U/2$, distinct in case of zero
bias, is diminished when the bias is finite. This maximum has been
associated with the equality of
$\mathcal{G}_{\uparrow}^{AP}=\mathcal{G}_{\downarrow}^{AP}$ and
subsequently $\langle n_{\uparrow}\rangle^{AP}=\langle
n_{\downarrow}\rangle^{AP}$ as discussed for $eV=0$. For finite
bias this relation is still fulfilled. It can be checked by
analyzing spin components of the differential conductance in
Coulomb blockade region.  However, the conductance for
antiparallel configuration is enhanced by finite bias and the
splitting and shifting of the conductance peaks also appears.  It
causes a gradual diminishing of TMR maximum. The negative TMR
minimum present for zero bias and associated with the rapid change
of the conductance polarization $P_{\mathcal{G}}^{AP}$, has
disappeared at finite bias. It has been shown in \cite{firstpaper}
that this sudden conductance polarization  switching also rapidly
diminishes when finite bias is applied.

\subsection{Large bias regime}
Consider the case of $eV=U$ for which the left (right) chemical
potential $\mu_{L}$ ($\mu_{R}$)is shifted upwards (downwards) by
$U$ on energy scale. The TMR vs. gate voltage for such a bias and
$\alpha=0.1$ is presented in Fig.~(\ref{fig5}) along with
corresponding differential conductances.
\begin{figure} [tr]
\epsfxsize=0.45\textwidth \epsfbox{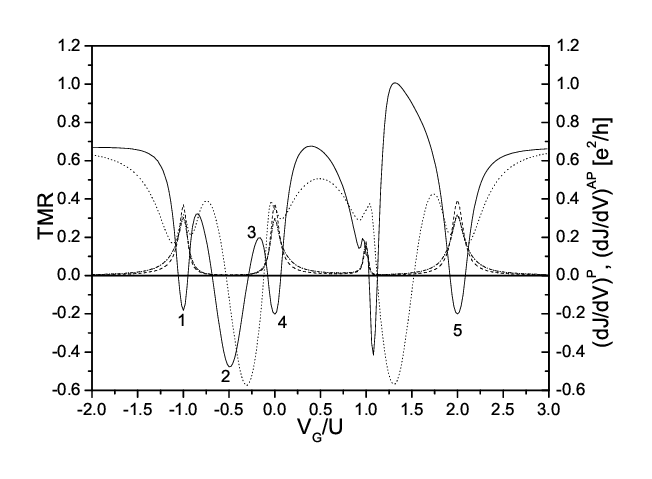}
\caption{\label{fig5}TMR dependence on gate voltage (solid curve)
for finite bias $eV=U$ calculated at $T=0.01U$, $P_{L}=0.5$ and
asymmetric coupling to the leads $\alpha=0.1$. The corresponding
differential conductances $(\partial J/\partial V)^{P}$
(dash-dotted curve) and $(\partial J/\partial V)^{AP}$ (dashed
curve) are also shown. The dotted TMR curve is for symmetric
coupling $\alpha=1$. }
\end{figure}
Let us discuss  various anomalies appearing in the TMR curve
labelled by arabic numbers. The TMR minima placed at the
differential conductance resonances have single particle origin.
For instance, in the point (1) at $V_{g}=-U$ the first hubbard
level $\epsilon_{d}^{I}$ coincides with $\mu_{L}$ and the second
$\epsilon_{d}^{II}$ lying above is empty. At this point TMR has a
(negative) minimum. This kind of minimum has appeared already for
zero and small bias each time when the QD hubbard level crossed
chemical potential of the leads: for $eV=0$ it corresponds to the
minimum at $V_{g}=0$ (see Fig.~(\ref{fig1})), and to the minimum
(1) for small bias (Fig.~(\ref{fig4})). Similar  correspondence is
for minimum (5) at $V_{g}=2U$ where the second hubbard level
$\epsilon_{d}^{II}$ coincides with $\mu_{R}$ and
$\epsilon_{d}^{I}$ is fully occupied. This minimum is analogous to
the minimum at $V_{g}=U$ for zero bias and minimum (4) for small
bias in Fig.~(\ref{fig4}). The minimum (4) at $V_{g}=0$ where
$\epsilon_{d}^{II}$ in resonance with $\mu_{L}$ corresponds to the
minimum (3) for small bias. One can also note series of maxima and
minima appearing in-between conductance resonances; for instance
minimum (2) and maximum (3). These anomalies are caused by
electron correlations and correspond to similar features of TMR at
zero bias (Fig.~(\ref{fig1})) at $V_{g}\sim U/2$. For large bias,
$eV\gg \Gamma_{\sigma}$, when the conductance resonances are well
separated in energy scale, the Coulomb blockade TMR anomalies can
be identified and their correspondence with zero bias TMR features
can also be established. In the small bias regime,
Fig.~(\ref{fig4}), they are considerably diminished.

The TMR minima of single particle origin, for instance those
numbered by (1), (4) and (5), transform into local maxima for
symmetric dot-leads coupling, $\alpha=1$ (dotted curve). Similar
feature was present for zero bias case, Fig.~(\ref{fig1}). Thus,
by experimentally tuning the coupling asymmetry one can resolve
the mechanism causing a given TMR anomaly. Note that the TMR curve
for $\alpha=1$ is symmetric with respect to $V_{g}=U/2$. At this
point the hubbard levels lay in the middle of the transport window
of the width $2U$. For instance, when $V_{g}$ is set to zero (to
U), the $\epsilon_{d}^{II}$ ($\epsilon_{d}^{I}$) comes into
resonance with $\mu_{L}$ ($\mu_{R}$) and the other hubbard level
is shifted to the center of the transport window. It gives the
same TMR feature for $\pm V_{g}$.
\begin{figure} [tr]
\epsfxsize=0.45\textwidth \epsfbox{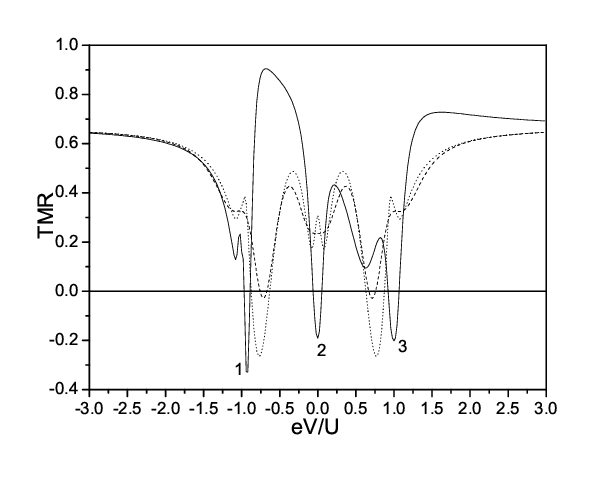}
\caption{\label{fig6}TMR dependence on applied bias for the set
gate voltage $V_{g}=0$ and $P_{L}=0.5$, calculated at $T=0.01U$,
and $\alpha=0.1$-solid curve and $\alpha=1$-dotted curve; dashed
curve- TMR for $\alpha=1$ and $T=0.06U$.}
\end{figure}

Consider now the bias dependence of TMR for the set gate voltage.
In Fig.~(\ref{fig6}) the TMR bias dependencies for $\alpha=0.1$
and $\alpha=1$ calculated for $V_{g}=0$ are shown. Again, the TMR
minima labelled by (1), (2) and (3), present for large coupling
asymmetry $\alpha=0.1$, can be identified as of single particle
origin. For minimum (2) at zero bias $\epsilon_{d}^{I}$ is in
resonance with both $\mu_{L}$ and $\mu_{R}$. For  minimum (1) at
$eV=-U$ the $\epsilon_{d}^{II}$ hubbard level is in resonance with
$\mu_{L}$ chemical potential and minimum (1) at $eV=U$ the
$\epsilon_{d}^{II}$ hubbard level is in resonance with $\mu_{R}$
chemical potential. These three minima disappear for symmetric
coupling $\alpha=1$ and the local maxima develop instead as
discussed before. The minima and maxima in the regions of
$eV\sim\pm U/2$ correspond to electron correlations induced TMR
anomalies discussed previously for zero bias, Fig.~(\ref{fig1}).
These anomalies appear close to particle-hole symmetric case, when
the Fermi level is situated in-between dot hubbard levels. In the
present case, at $eV=-U/2$ the $\mu_{L}$ chemical potential is
placed in-between hubbard levels in the sequence:
$\mu_{R}<\epsilon_{d}^{I}<\mu_{L}<\epsilon_{d}^{II}$. At $eV=U/2$
the $\mu_{R}$ is in-between hubbard levels in sequence:
$\mu_{L}<\epsilon_{d}^{I}<\mu_{R}<\epsilon_{d}^{II}$ in energy
scale. From comparison of the TMR curves for $\alpha=1$ and
$T=0.01U$ and at a higher temperature $T=0.06U$ one notes that the
minima caused by electron correlations are much more sensitive to
the temperature increase than maxima, similarly as in the case of
zero bias, Fig.~({\ref{fig3}). Thus, by the increase of
temperature one can distinguish the TMR minima of single particle
origin which are robust to the temperature change from those
caused by electron interactions, sensitive to temperature.

For symmetric coupling to the leads the TMR curve becomes
symmetric with respect to $eV=0$. It is understood when one
notices that the change of the bias direction is equivalent to the
simultaneous exchange of $\mu_{L}\leftrightarrow\mu_{R}$ and
$\epsilon_{d}^{I}\leftrightarrow\epsilon_{d}^{II}$. It gives for
symmetric coupling, $\alpha=1$, the same value of TMR for $\pm
eV$.
\section{Influence of effective magnetic fields}
\begin{figure} [tr]
\epsfxsize=0.45\textwidth \epsfbox{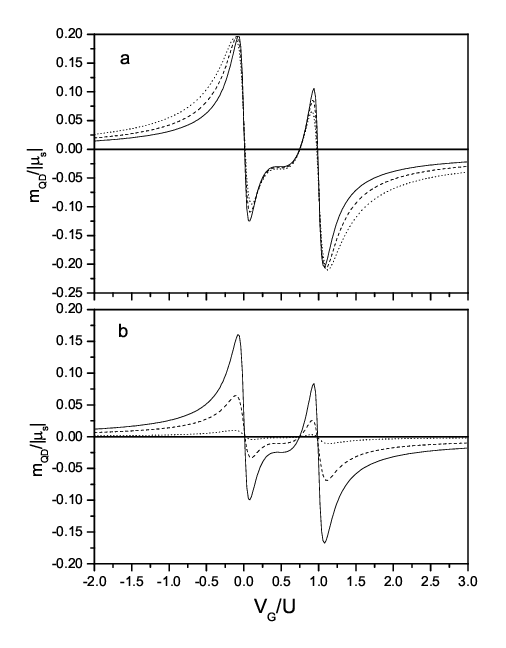} \caption{\label{fig7}
Dot magnetic moment vs. gate voltage for parallel (a) and
antiparallel (b) configuration for various asymmetry $\alpha$.
Panel (a) and (b): $\alpha=0.1$ solid line, $\alpha=0.5$-dashed.
Dotted line in panel (a) for $\alpha=1$ and in panel (b) for
$\alpha=0.9$ ($m_{QD}=0$ for $\alpha=1$ in antiparallel
configuration). Calculated for zero bias at $T=0.01U$,
$P_{L}=0.5$.}
\end{figure}
The dot attached to spin-polarized leads is influenced by two
effective magnetic fields, which have different origin.  i) The
hopping of electrons from the leads, which have an excess of one
of the spin components, produces an occupancy polarization of the
dot. Thus, the dot acquires magnetic moment
$m^{\beta}_{QD}=(\langle n_{\uparrow}\rangle^{\beta}-\langle
n_{\downarrow}\rangle^{\beta})|\mu_{s}|$, where
$\mu_{s}=g\mu_{B}s_{z}$ is the magnetic moment associated with
spin-dependent (but degenerate) sub-levels of $\epsilon_{d}$ and
$s_{z}=\pm 1/2$. This effect can be regarded as if the dot were
under influence of an external magnetic field and the Zeeman
splitting  of the dot energy level is produced. However, the value
of the resultant dot's magnetic moment at given temperature is
regulated by the relative position of the dot level with respect
to the leads chemical potentials and electron-electron
interactions. It has non-monotonic behavior as shown in
Fig.~(\ref{fig7}). It is in contrast to the usual Zeeman splitting
of the level by external field $H_{ext}$, for which the magnetic
moment $m=\tanh(|\mu_{s}|H_{ext}/k_{B}T)$ increases monotonically
with increase of the field. Let us discuss the general features of
the dot magnetic moment for parallel and anti-parallel
configuration, Fig.~(\ref{fig7}). Because $P_{L}>0$ the spin
sub-levels of each dot hubbard level
$\epsilon_{d\uparrow}^{\gamma}$ and
$\epsilon_{d\downarrow}^{\gamma}$ ($\gamma=I,II$) have different
widths $\Gamma_{\uparrow}>\Gamma_{\downarrow}$. For gate voltage
large and negative the $\epsilon_{d}^{II}$ is empty and both the
$\epsilon_{d\uparrow}^{I}$ and $\epsilon_{d\downarrow}^{I}$ are
barely populated; the magnetic moment is small. The value of
$m_{QD}$ is positive because
$\Gamma_{\uparrow}>\Gamma_{\downarrow}$ and
$\epsilon_{d\uparrow}^{I}$ is being populated earlier than
$\epsilon_{d\downarrow}^{I}$. When $V_{g}$ increases, $m_{QD}$
also increases reaching the maximum. Further increase of $V_{g}$
causes faster filling of $\epsilon_{d\downarrow}^{I}$ sub-level
because it is sharper than $\epsilon_{d\uparrow}^{I}$. Thus,
$m_{QD}$ decreases towards zero at $V_{g}=0$, where
$\epsilon_{\sigma}^{I}$ is in resonance with Fermi energy
$\epsilon_{F}$. Further shift of $\epsilon_{\sigma}^{I}$ causes
further increase of $\langle n_{\downarrow}\rangle$ until a
minimum of $m_{QD}$ is reached. Note, that the sequence of the
filling of the first hubbard level $\epsilon_{\sigma}^{I}$, caused
by inequality of the $\Gamma_{\uparrow}$ and
$\Gamma_{\downarrow}$, is additionally enhanced by
electron-electron interactions, manifested by the spectral weight
dependence $\sim(1-\langle n_{\bar{\sigma}}\rangle)$ of
$\epsilon_{\sigma}^{I}$, Eq.~(\ref{G1gen}). After reaching of
$m_{QD}$ minimum, the second hubbard level
$\epsilon_{\sigma}^{II}$ starts to be filled and $m_{QD}$
increases towards zero due to the same mechanism
($\Gamma_{\uparrow}>\Gamma_{\downarrow}$) as in the region of
$V_{g}\sim 0$. The $m_{QD}=0$ at Coulomb blockade corresponds to
the particle-hole symmetry case of $\epsilon_{d}=-U/2$ and
$\langle n_{\uparrow}\rangle=\langle n_{\downarrow}\rangle=0.5$
for unpolarized leads. Further increase of $V_{g}$ causes a
maximum of $m_{QD}$ to appear, followed by $m_{QD}=0$, when
$\epsilon_{d}^{II}$ is in resonance with $\epsilon_{F}$ for
$V_{g}=U$, and then minimum. Note that the $m_{QD}$ maximum is
weakened, as compared to one at $V_{g}\sim 0$, and the minimum is
enhanced. It is due to electron interactions: the spectral weight
of the second hubbard level $\epsilon^{II}_{\sigma}$ is $\sim
\langle n_{\bar{\sigma}}\rangle$ (see Eq.~(\ref{G1gen})). It also
causes that the dot magnetic moment remains negative for large
positive $V_{g}$ (compare to $m_{QD}>0$ for large negative
$V_{g}$).

An increase of the $\alpha$ symmetry  of the dot-leads coupling
causes a gradual decrease of the dot magnetic moment for (AP)
configuration, panel (b). For symmetric case of $\alpha=1$ the
magnetic moment is zero in whole range of gate voltages as has
been pointed out in the discussion of TMR maximum at zero bias.
For $\alpha < 1$ the $m_{QD}=0$ line collapses into one $V_{g}$
point at which the TMR maximum appears.

ii) The second field is an effective magnetic field, $H_{eff}$,
produced by magnetized electrodes. Its value is not dependent on
gate voltage but rather on the relative leads polarizations. While
TMR measurement is performed, the external magnetic field is
applied to the system, which changes relative leads polarizations.
Due to the intentionally different shape anisotropy of the left
and right lead, the magnetization of each lead responds
differently to the external field. It enables  anti-parallel lead
polarization for small field and  parallel configuration for
higher field. The dot itself is in turn subjected to an effective
magnetic field produced by the magnetic electrodes nearby. This
field is the largest for parallel configuration.

For InAs quantum dots gyromagnetic ratio is negative, $g<0$
\cite{doty}, thus the spin moments of $\epsilon_{d\uparrow}$ and
$\epsilon_{d\downarrow}$ point in the same direction as the
corresponding magnetic moments. The dot level subjected to the
field $H_{eff}$ produced by the polarized leads (of direction
opposite to the leads spin polarization) is Zeeman
split:$\epsilon_{d\downarrow}=\epsilon_{d}-\Delta$ and
$\epsilon_{d\uparrow}=\epsilon_{d}+\Delta$,
$\Delta=|g\mu_{B}H_{eff}/2|$.  The Zeeman energy splitting for
$|g|=3.8$ \cite{igarashi} gives $2\Delta= 0.22 meV/T$ ($= 0.015 U$
per tesla). We considered the influence of $H_{eff}$ on the TMR
features, assuming $H_{eff}^{P}$ of the order of $\pm 0.1 T$,
taken from experiments \cite{hamaya1,hamaya2}. For antiparallel
configuration the cancellation of the fields originating from the
leads has been assumed $H_{eff}^{AP}=0$. We have found that the
possible Zeeman splitting of the dot level has negligible effect
on TMR in this field range.
\section{Conclusions}
We have discussed the TMR anomalies encountered for the quantum
dot coupled to spin-polarized leads in the regime of Coulomb
blockade. We have shown that there are two kinds of such
anomalies. One kind has the single particle origin  and can be
interpreted in the frame of non interacting electrons model. The
second kind of anomalies is caused by electron interactions. The
TMR minima (and its sign change) of single particle origin  appear
at the conductance resonances for asymmetric dot-lead coupling.
They are robust to the temperature increase and gradually
transform into local maxima when the symmetry of the dot-lead
coupling increases. The anomalies associated with electron
interactions appear at Coulomb blockade, in-between conductance
resonances. The TMR maximum at Coulomb blockade, far exceeding 100
\% is of this origin. It appears when in antiparallel
configuration both the current the dot occupancy are spin
unpolarized. This maximum survives at typical temperatures of
experiment. We also predict the TMR sign change at Coulomb
blockade. It appears due to the rapid polarization switching of
the current in (AP) configuration and the enhancement of the
conductance in one of the (AP) spin channels by the dot-leads
coupling asymmetry. It is very sensitive to the increase of
temperature and depends on the initial polarization of the current
coming from the leads. We have shown that the nature of the
discussed anomalies can be experimentally resolved by the change
of the dot-leads coupling asymmetry and/or temperature. Finally,
we have analyzed the dot polarization, as induced by the coupling
to the polarized leads, and shown that it also depends on electron
interactions present inside the dot. The estimated Zeeman field
splitting, produced by the leads, has negligible effect on TMR for
experimental range of the fields.

\begin{acknowledgements}
The work is supported from the European Science Foundation
EUROCORES Programme FoNE by funds from the Ministry of Science and
Higher Education and EC 6FP (contract N. ERAS-CT-2003-980409).
\end{acknowledgements}

\end{document}